\documentstyle[12pt,aasms4,amssym]{article}

\received{12 May 1998}

\slugcomment{Submitted to the Astrophysical Journal Letters.}

\lefthead{M. Georganopoulos \& A.P. Marscher}
\righthead{The Blazar Inner Jet}

\begin{document}

\title{Modeling the variability of the BL Lacertae  object PKS 2155-304.}

\author{Markos Georganopoulos and Alan P. Marscher}
\affil{Astronomy Department, Boston University, 725 Commonwealth Avenue, Boston,  MA 02215}

\begin{abstract}

The bright X-ray selected BL Lacertae object   PKS 2155--304 has been the 
target
of two intense multiwavelength campaigns,  in  November 1991 and in May 1994.
 Although
the spectral energy distributions  at both epochs were quite similar, 
the source 
exhibited two very distinct  variability patterns  that cannot be easily 
 reconciled with homogeneous, one--zone jet models. During the first epoch
the variability was almost achromatic in amplitude, 
with a time lag between X-rays and UV
of $\approx 3$ h, while during 
the second epoch the variability amplitude increased as a function of 
wavelength, 
with the EUV flare peaking $\approx 1$ day after the X-ray flare. 
We model the source using a time-dependent 
inhomogeneous accelerating jet model. We reproduce the general characteristics
of the different variability
 signatures by 
  assuming that  plasma disturbances with different physical properties 
propagate
downstream in an underlying  jet characterized by the same set of 
physical parameters
at both epochs. 
A time delay of $\approx$ 1 day between the hardening
 of the UV spectral index and the UV flux, present at  both epochs,
is modeled with stochastic fluctuations in the particle acceleration 
manifested through  small  variations of the  maximum energy of 
the injected electrons.  
We predict that  similar time delays will  be present in future observations,
even in the absence of strong variability events.
We stress the importance of observations at  neighboring frequencies
as a diagnostic tool for the structure of the quiescent jet in blazars, 
especially in the seemingly dull  case when  strong variability is absent.

\end{abstract}


\pagebreak

\section{INTRODUCTION}

The X-ray selected BL Lacertae object (XBL) PKS 2155-304 is a nearby (z=0.116)
 bright blazar. 
 The source has been monitored intensively twice in a frequency 
range from
radio waves to X-rays.
The first campaign took place during November 1991 (\cite{edelson95}).  
Observations were made at UV 
 (\cite{urry93}), X-rays  (\cite{brinkmann94}), and radio, infrared,  
and optical wavelengths  (\cite{courvoisier95}).
There was a flux
increase by a factor of $\approx 2$ during the 30--day period of the campaign
in the IR to X-ray frequency regime. Superposed on  this slow variation were 
flares  with an amplitude of $\approx 10 \% $ and duration of 
$\approx 0.7$ days.
 The fractional amplitude of the variations appeared almost constant in the energy 
regime from IR to X-rays.  The X-ray variations led the UV by 
 $\approx 3 $ h, while no delay down to a limit of $\approx 2$ h
was found between the UV and the optical.
These  characteristics are  difficult to reconcile with 
homogeneous  synchrotron source  models.

During the  May 1994 campaign
observations were made at X-ray (\cite{kii98}), 
extreme UV (\cite{marshall98}), 
UV (\cite{pian97}),
and optical,  IR, and radio wavelengths (\cite{pesce97}). 
The results are summarized
by Urry et al. (1997). The ASCA observations recorded an X-ray flare on May 19.
The flare was symmetric with a duration of $\approx 0.8 $ days, and a 
relative amplitude of $\approx 2$ . The hard X-rays
(2.2-8 KeV) led the soft X-rays (0.5-1 KeV) by $\approx 5000$ s 
(\cite{kii98}). EUVE recorded a flare that started on May 19 and lasted for
$\approx 1.5$ days with a relative amplitude of $\approx 50 \%$ 
(\cite{marshall98}). 
Cross--correlation analysis showed that the EUVE flare 
lagged the X-ray flare by $\approx 1$ day. Finally, IUE observations 
revealed  a UV flare that peaked $\approx 2 $ days after the X-ray flare
with a relative amplitude of $\approx 35 \%$ (\cite{pian97}). 
The IUE flare lasted for
$\approx 2.5 $ days. 
Although the variability behavior during these two campaigns was quite 
different, the spectral energy distributions (SEDs) at  both epochs were  very
similar. 
This encourages us to identify the SED with the radiation produced 
by the underlying
undisturbed jet and the variable emission with radiation  from newly
injected plasma components. The different variability signatures can then
be  interpreted simply as manifestations of different physical properties of the various plasma disturbances
propagating along the jet.

\section{THE MODEL}

We use the accelerating inner jet model of Georganopoulos \& Marscher (1998),
upgraded  to take into account the time evolution of the observed radiation
due to a propagating disturbance in the plasma flow.
The jet is characterized by
the Lorentz factor $\Gamma_{\star}$ of the bulk motion of the injected plasma
at the base of the jet, 
the distance $z_{\star}$ of the base of the jet from the  stagnation point of 
the flow,
the radius $r_{\star}$ of the base of the jet, and
the exponent $\epsilon$ that describes how fast the jet opens and accelelerates
($r\propto z^{\epsilon}$, $\Gamma\propto z^{\epsilon}$). The injected plasma 
has an electron kinetic luminosity 
$\Lambda_{kin}$ and is characterized by a power law electron energy distribution (EED),
$N(\gamma,z_{\star})=N_{\star} \gamma^{-s}$,  
$ \gamma_{min}\leq\gamma\leq\gamma_{max} $. 
The comoving magnetic field, assumed predominantly tangled with a small component 
aligned to the jet axis, decays 
according to the relation $B\propto r^{-1}$. 
The electrons lose 
energy due to adiabatic and synchrotron losses, and the EED evolves as the
plasma flows downstream, giving rise to a local  non-power law synchrotron
spectrum that evolves along the jet axis. This formulation does not include 
Compton losses, but in the case of PKS 2155--304 these are probably not
the  dominant energy loss mechanism (Vestrand et al. 1995).

The modeling of relativistically moving disturbances requires inclusion of
  light--travel time delays and light aberration, which have not been 
considered in previous calculations (\cite{celotti91}). If a disturbance travels with a speed $\beta c$ in a 
direction
that forms an angle $\theta$ with the line of sight, the
observed transverse velocity is 
\begin{eqnarray}
v=\frac{\beta c\sin\theta}{(1-\beta \cos\theta)(1+z)} \nonumber,
\end{eqnarray}
where z is the redshift of the source.
This creates a combination of an apparent rotation and distortion of the 
disturbance front. 
We calculate the position of the disturbance as a function of time in 
the observer's frame,
taking into account relativistic time delays, and then  perform the 
radiative transfer, taking into account the effect of  light
aberration (our treatment is similar to that of G\'{o}mez et al. 1994).

\begin{figure}
\epsscale{0.8}
\plotone{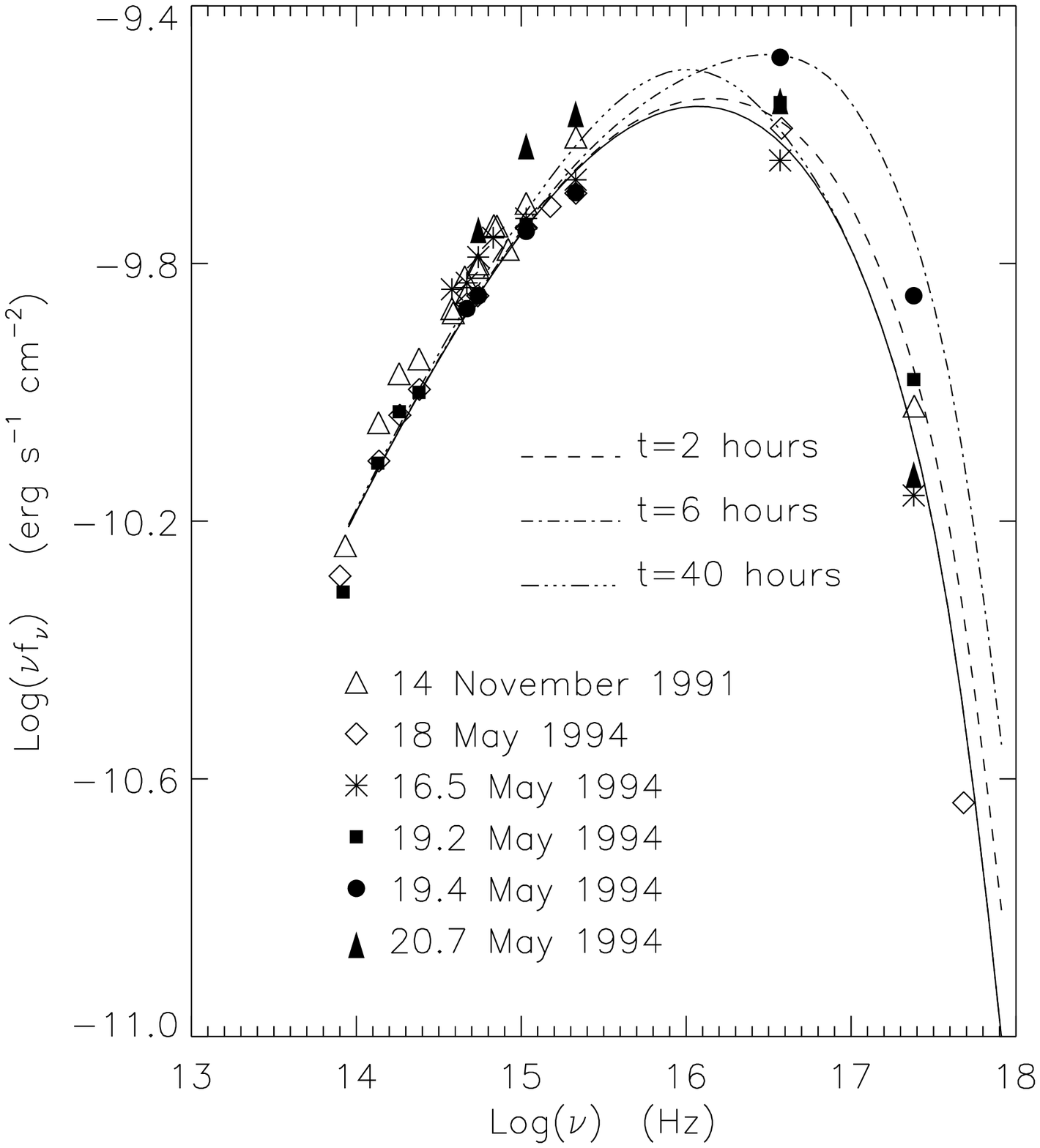}
\caption{The spectral energy distribution (SED) of PKS 2155-304 for 14 November 1991 (triangles; 
Edelson et al. 1995),  and 19 May 1994 (diamonds; Urry et al. 
1997). The data  of 16.5 May 1994 (asterisks) were selected  by Urry et al. (1997)  to represent the pre-flare SED. The filled symbols represent the observed time evolution of the SED during
the 19 May 1994 flare. The solid curve represents the steady--state model SED. The broken 
curves represent the time evolution of
the simulated flux during the May 94 flare. The flare starts at time t=0 
in the 
observer's frame, which corresponds to $\approx 19.1 $ May.  The values  of the steady--state model parameters are: 
\protect $r_{\star}=10^{15}$ cm, $z_{\star}=1.3 \;10^{15}$ cm, 
$\Gamma_{\star}=1.5$, $\epsilon=0.3$, $B_{\star}=0.1$ G, $\Lambda_{kin}=10^{46}$ erg/s, $\gamma_{max}=7.5 \; 10^{5}$, $s=1.7$, and $\theta=14^{\circ}$. }
\label{steady}
\end{figure}

 Modeling the steady-state, quiescent  emission of 
 a blazar requires a set of simultaneous
multiwavelength data that correspond to the low state of the source.
 An 
extensive period of minor activity or the time  before a significant event
can be used as working definition of the low state of a jet.
We construct (see Fig. \ref{steady}) low state SEDs of 
PKS 2155-304 using multiwavelength data taken in 1991 November 14, during a
 period of 18 h (see \cite{edelson95}), and in 1994 May 18, 
during a period of $\approx 1$ day (see \cite{urry97}), when local flux minima,
followed by a flare, were observed.
The two SED appear quite similar, and, taking into account the problems of 
simultaneity and of  definition of the quiescent state of the  source, we 
 assume that
the undisturbed plasma flow at  both epochs can be 
described adequately  by one single set
of physical parameters.  We fit this composite two--epoch SED using the
model parameters  given in the caption of Figure \ref{steady}.
The solid curve in Figure \ref{steady} shows the 
steady--state model SED.

\section{THE CAMPAIGN OF NOVEMBER 1991}

We focus  here on the  short time scale 
( $\approx 0.7$ days) 
 variability, characterized by achromatic peak fractional amplitudes and a 
 $\approx 3 $ h time delay
between  X-ray and  UV energies. We interpret the small 
time delays as an indication of near co--spatiality of the regions
emitting the variable components at different frequencies. This can be achieved
if the injected plasma component is characterized by a high magnetic field,
so that the radiative loss time scale for the electrons emitting
the X-ray photons is of the order of the observed time lag of $\approx 3 $ h 
between the X-ray and the UV light curve. The fact that the 
doubling time scale inferred by the small amplitude ($\approx 10 \% $) and
the $\approx 0.7$ days flare duration, is longer than the observed 3 h 
time delay, implies that the  duration of the flare
is not due to  radiative losses, but rather to the injection mechanism.
This further implies that
the size  of the injected component is  larger that the 
region that radiates at each frequency. As a result, the 
number of 
electrons that radiate at a given  frequency band is a function of frequency in
a manner similar to that of the underlying jet. This gives rise to a 
frequency independent fractional  
amplitude for the flare, as long as the spectral index
of the variable component  is similar to that of the underlying jet. 

\begin{figure}
\epsscale{0.8}
\plotone{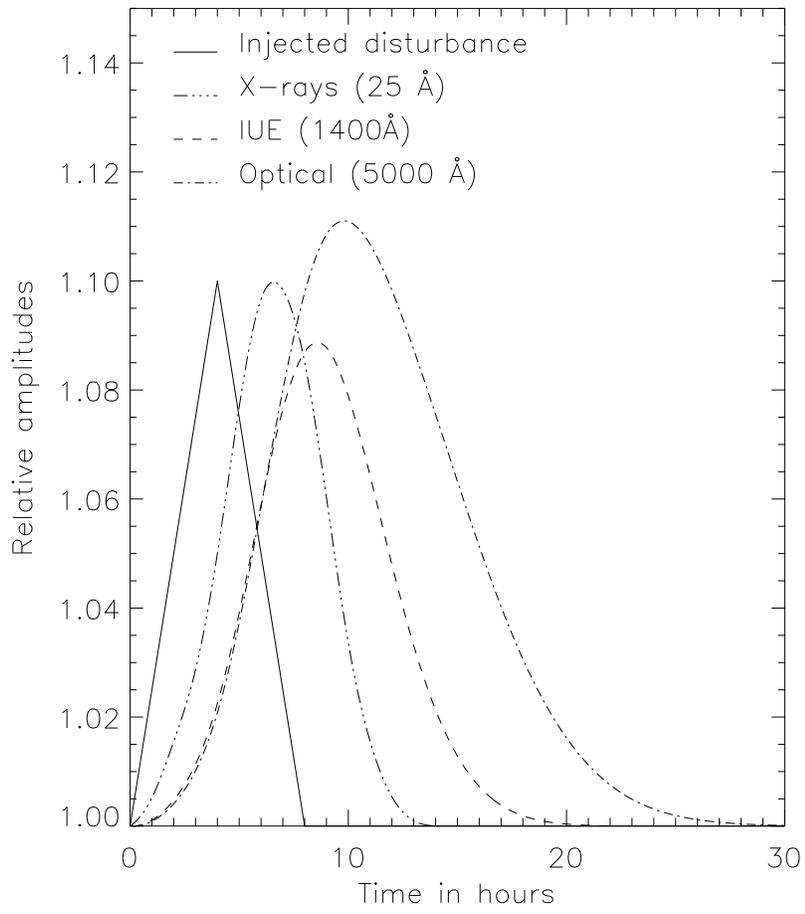}
\caption{Simulated light curves for the November 91 campaign.}
\label{91}
\end{figure}

We assume that a ``slug'' of plasma  characterized by a magnetic field 
$B_{inj}=1.2$  G 
was injected at the base of the jet. The injected electron kinetic
luminosity  increased linearly with time to reach a maximum amplitude of 
$\Lambda_{inj}=0.1 \Lambda_{kin}$  after four hours (solid curve
 in Figure \ref{91}),
and then declined linearly over four more hours. 
The shape and cutoff energies are assumed to have been the same as for the 
quiescent jet.
The response of 
the jet is also
shown in terms of the relative amplitude of the light curves at X-ray, UV, 
and optical energies. Figure \ref{91} should be compared to Fig. 2 of 
Edelson et al. (1995). The time delays are in agreement with the observations,
and the fractional amplitude of variability, although not precisely
 achromatic, 
 does not vary significantly with frequency.

\section{THE CAMPAIGN OF MAY 1994}

The clear temporal separation between the peak flux at different
frequencies translates into a spatial separation. Combining this with the
assumption that a single plasma component  propagating downstream was responsible 
for the observed variability, we again assume that
a component of high energy electrons is injected at the base of the jet. 
As before, these
electrons emitted  synchrotron radiation at high frequencies (X-rays) 
and cooled as they propagated downstream, radiating at progressively lower 
frequencies. This can explain the time delays between the X-rays
and the EUV and UV energies, but the longer time lags at the lower frequencies
require the injected electron population to have beem more monoenergetic than in November 1991, i.e., $\gamma_{min} \rightarrow \gamma_{max}$. 
The  relative amplitude of the flare
was frequency dependent, with the higher frequencies
characterised by a  higher amplitude of variability (\cite{urry97}). 
Given the fact that the number of 
electrons that contribute to the underlying jet emission increases as the 
frequency decreases, the fixed number of electrons in the plasma component will
give rise to a flare of higher relative amplitude  at higher frequencies. 

\begin{figure}
\epsscale{0.8}
\plotone{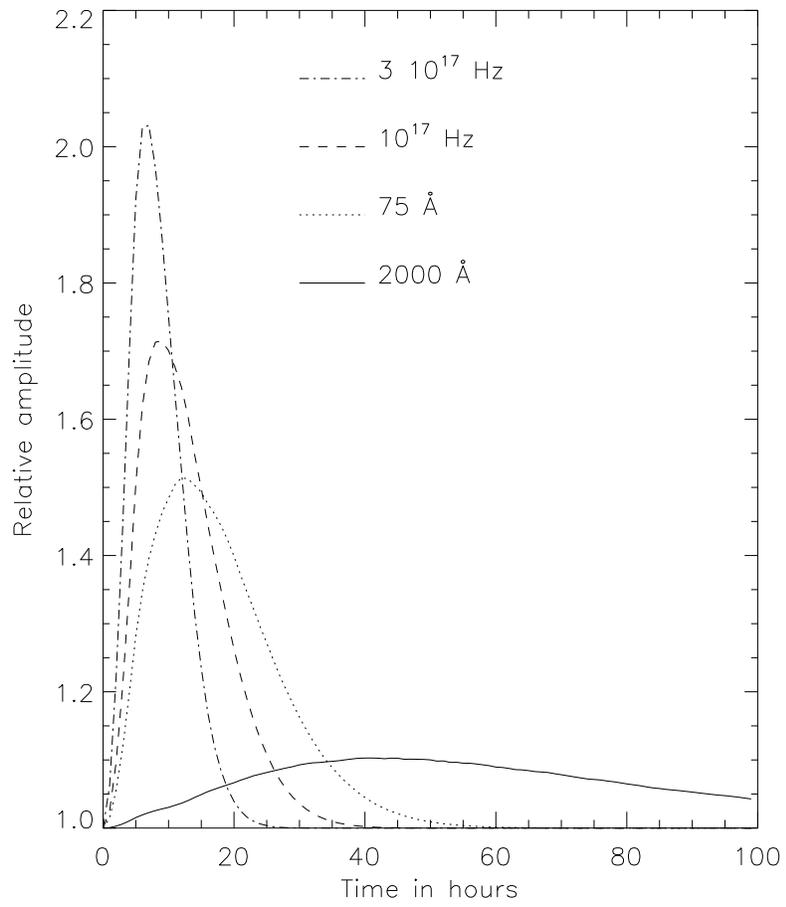}
\caption{Simulated light curves for the May 94 campaign.}
\label{94}
\end{figure}

We therefore model the May 1994 flare with an injected  EED
 confined by $2\times 10^5 \leq \gamma \leq 7.5 \times 10^{5}$, with $N(\gamma)
\propto \gamma^{-s}$, $s=2.1$, and  injected luminosity 
$\Lambda_{inj}=1.3 \;\Lambda_{kin}$. We adopt for  the duration of the 
injection event  $t_{inj}=4$ h in the observer's frame.
 In Figure \ref{94} we plot the 
simulated light curves at  different frequencies.  In Figure \ref{steady} we
plot  the observed
SED before and through the May 19 flare (the same data are plotted in
Fig. 1 of Urry et al. 1997). The relative amplitudes and the
time delays are in good agreement with the observations, with the exception 
of the amplitude of the UV flare, which had a relative amplitude of $35 \%$, 
while the simulation gives an amplitude of $10 \%$. An  explanation 
for this  discrepancy  could be a mild 
re-acceleration of the electrons in the injected plasma component. 
Such a mild re-acceleration would be manifested mainly at  lower frequencies,
thereby increasing  the flare amplitude relative to that at
higher frequencies. 
An interesting feature is the $\approx 1$ hour  
time delay between the $3 \times 10^{17}$ Hz  and the  $10^{17}$ Hz X-rays, 
as well as the smaller
amplitude of the lower energy variation. 

\begin{figure}
\epsscale{0.8}
\plotone{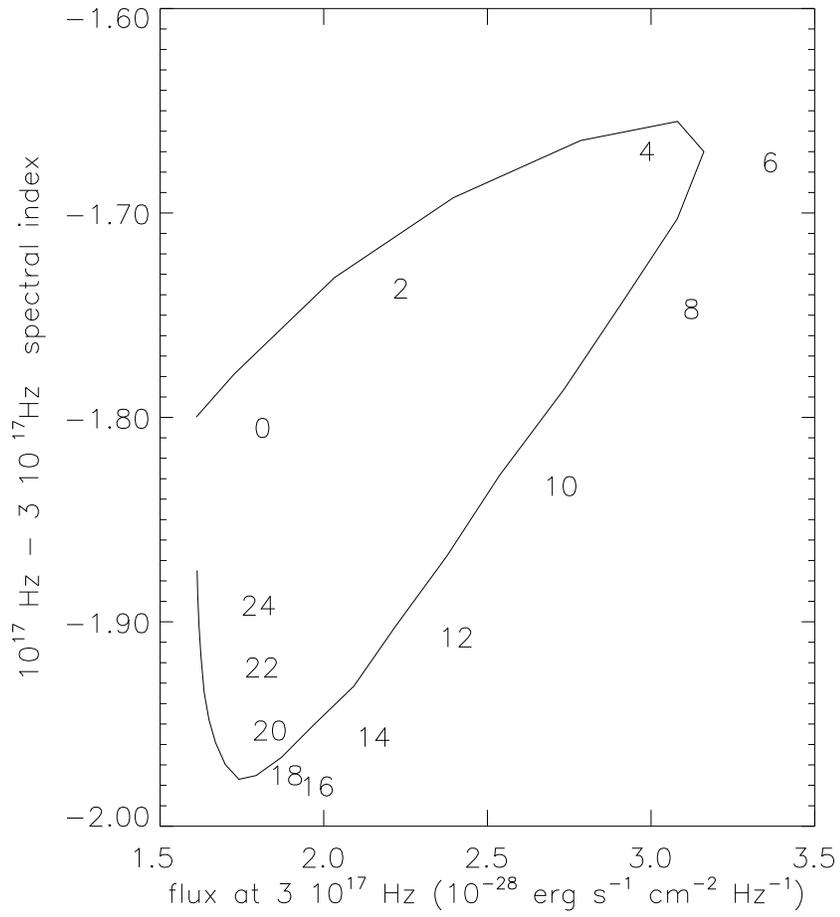}
\caption{Model X-ray spectral index vs X-ray flux for the May 1994 flare. The integers following the clockwise loop indicate time in hours.}
\label{clock}
\end{figure}

In Figure \ref{clock} we plot the
X-ray spectral index versus X-ray flux diagram for the simulated flare. The
amplitude of the X-ray spectral index variation ($\Delta \alpha \approx 0.3$)
is slightly smaller than observed  ($\Delta \alpha \approx 0.5$)
We obtain the familiar clockwise  loop that has been observed in
X-ray variability studies
of other sources (e.g. see \cite{takahashi96} for a similar curve in Mkn 421). Such a 
behavior is not universal
and should be expected only for isolated flares 
that do not blend with any preceding  or subsequent variability.

\section{THE SPECTRAL INDEX -- LUMINOSITY LAG}
The results of cross--correlation 
analysis  between the UV flux and spectral index, both for 
the November 91 flare (\cite{urry93}) and the May 94 flare (\cite{pian97}), 
show that the variation in the UV spectral index leads  the change  in intensity by 
$\approx 1$ day, in the sense that an increase (decrease) in intensity
is preceded by a hardening (softening) of the spectral index.
Given the very disparate variability behavior observed in the two campaigns,
this similarity is not intuitively expected. Particularly puzzling at first
look is the fact that the time delay persists in the May 94 campaign even
after the central flare has been excluded from the cross--correlation analysis.

\begin{figure}
\epsscale{0.8}
\plotone{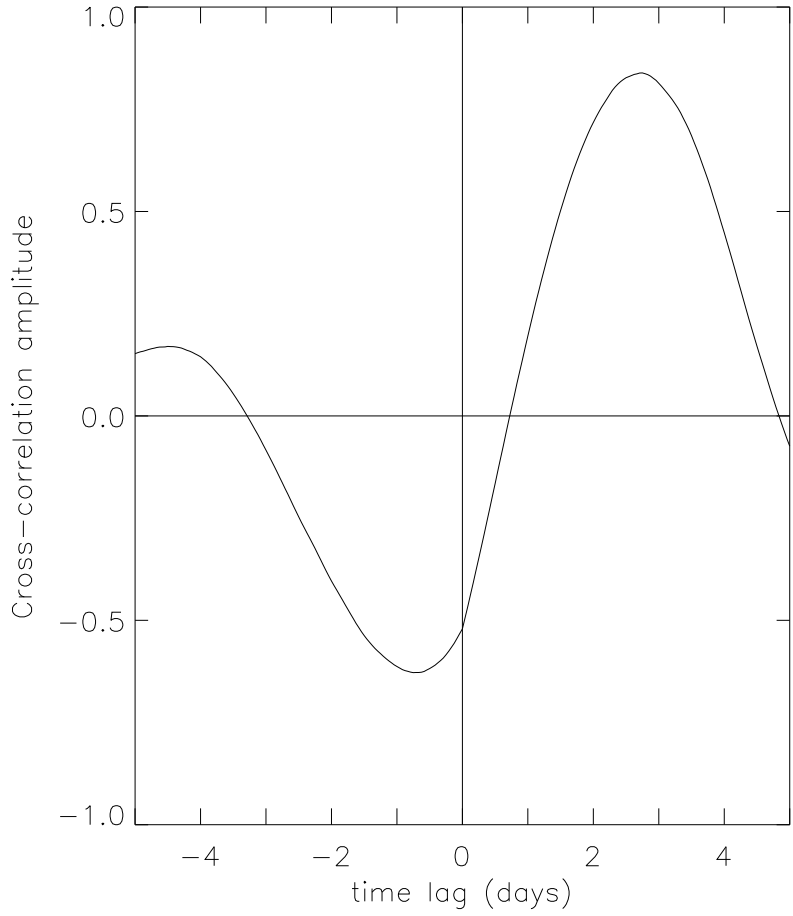}
\caption{Cross--correlation function between the model flux at 2000 \AA \ and the        1400--2800 \AA \ spectral index. The spectrum flattens (steepens) 
$\approx 1$ day before the flux increases (decreases).}
\label{crosscor}
\end{figure}

We propose that such a behavior is the response of the jet to small 
variations of the upper cutoff $\gamma_{max}$ of the injected EED. 
It is 
plausible that such stochastic fluctuations are always  present, regardless
of the existence or absence  of major 
newly injected components that produce the
large amplitude variability.
Such variations are equivalent to the addition or subtraction of 
a small amplitude, nearly monoenergetic electron  component that propagates 
downstream, affecting first the higher and then the lower frequencies.
At neighboring frequencies this results in 
two low amplitude flares with the higher frequency flux peaking
first. In Figure \ref{crosscor} we plot the cross--correlation function
between the  flux at 2000 \AA \ and the  1400--2800 \AA \ spectral index,
when the upper cutoff $\gamma_{max}$ of the injected EED increases from
$\gamma_{max}=7.5 \times 10^5$ to $\gamma_{max}=7.6 \times 10^5$ for eight hours,
while the rate of injected particles is held constant.
Although the variation amplitude is small ($\approx$ 1 \%), the lag
of $\approx 1$ day is the same as that observed 
(cf. Fig. \ref{crosscor} to Fig. 7 of Pian et al. 1997 and 
 Fig. 9 of Urry et al. 1993).

Since this time lag is the response of the jet to stochastic variations 
of $\gamma_{max}$, we predict that the lag  will  be present in future 
observations
of the source, particularly during  intervals of mild activity
when no major variability events occur, although the exact magnitude of 
the delay will depend on  the nature of the variations.
Similar considerations are  relevant for all  blazars.  
Observations at different 
frequencies will in principle  produce time lags of different magnitudes.
This ``spectral fluctuation'' technique  can be used to 
constrain  the geometry and physics of the quiescent jet, as well as provide
information about  the  power spectrum of  fluctuations  in the density
and energy distribution of the injected electrons.
The method does not rely on the sporadic
appearance of  strong variability events, which in any case  are related more
to the newly injected components and less to the underlying jet flow. 
Mild variability of the order of a few percent is sufficient for the 
cross--correlation analysis. 

This research was supported in part by  NASA Astrophysical Theory Program
grant NAG5-3839.

\clearpage

\end{document}